\documentstyle[aps,eqsecnum,amsfonts]{revtex}

\input epsf

\def\R{\mbox{$\Bbb R$}}     
 
\newcommand{\e}{{\rm e}}
\newcommand{\grad}{{\rm grad}}
\newcommand{\half}{\frac{1}{2}}

\newcommand{\calm}{{\cal M}}
\newcommand{\calv}{{\cal V}}
\newcommand{\pone}{{\cal L}_k^\pm}
\newcommand{\ptwo}{{\cal K}_{\cal M}}
\newcommand{\pthree}{P_{\cal S}(I)}
\newcommand{\pfour}{P_{\cal S}^\pm(VII_h)}
\newcommand{\pfive}{P(I)}
\newcommand{\psix}{P(VI_h)}
\newcommand{\pseven}{{\cal F}_{\cal S}(I)}
\newcommand{\peight}{{\cal A}_{\cal S}(VI_h)}
\def\q{\quad}

\begin{document}
  
\title{ Scalar Field Cosmologies with Barotropic Matter: Models of Bianchi 
class B}

\author{A. P. Billyard, A. A. Coley}

\address{   Department of Mathematics and Statistics \\
		Dalhousie University, 
		Halifax, Nova Scotia B3H 3J5}

\author{          R. J.  van den Hoogen}

\address{   Department of Mathematics, Statistics 
				and Computer Science,\\
		Saint Francis Xavier University, 
		Antigonish, N.S., B2G 2W5}

\author{           J. Ib\'a\~nez, I. Olasagasti}

\address{   Departmento de Fisica Teorica, 
		Universidad del Pais Vasco 
		Bilbao, Spain}

\maketitle

\begin{abstract} We investigate in detail the qualitative behaviour of the 
class of Bianchi type B spatially homogeneous cosmological models in
which the matter content is composed of two non--interacting
components; the first component is described by a barotropic fluid
having a gamma-law equation of state, whilst the second is a
non--interacting scalar field $\phi$ with an exponential potential
$V(\phi)=\Lambda\e^{k\phi}$.  In particular, we study the asymptotic
properties of the models both at early and late times, paying
particular attention on whether the models isotropize (and inflate) to
the future, and we discuss the genericity of the cosmological scaling
solutions.
\end{abstract}



 \section{Introduction}

Scalar field cosmology is of importance in the study of the early
Universe and particularly in the investigation of inflation (during
which the universe undergoes a period of accelerated expansion \cite{R1,Olive}).
One particular class of inflationary cosmological models are those
with a scalar field and an exponential potential of the form
$V(\phi)=\Lambda\e^{k\phi}$, where $\Lambda$ and $k$ are non-negative
constants.  Models with an exponential scalar field potential arise
naturally in alternative theories of gravity, such as, for example,
scalar-tensor theories.

Scalar-tensor theories of gravitation, in which gravity is mediated by
a long-range scalar field in addition to the usual tensor fields
present in Einstein's theory, are perhaps the most natural
alternatives to general relativity (GR).  In the simplest Brans-Dicke
theory of gravity (BDT; \cite{BD}), a scalar field,
$\phi$, with a constant coupling parameter $\omega_0$, acts as the
source for the gravitational coupling. More general scalar-tensor
theories have a non-constant parameter, $\omega(\phi)$, and a non-zero
self-interaction scalar potential, $V(\phi)$.  Observational limits on
the present value of $\omega_0$ need not constrain the value of
$\omega$ at early times in more general scalar-tensor theories (than
BDT).  Hence, more recently there has been greater focus on the early
Universe predictions of scalar-tensor theories of gravity, with
particular emphasis on cosmological models in which the scalar field
acts as a source for inflation \cite{Olive,T1}. BDT ( and other theories of
gravity, such as, for example, more general scalar-tensor theories and
quadratic Lagrangian theories and also theories undergoing dimensional
reduction to an effective four-dimensional theory \cite{R3}), are known to
be conformally equivalent to general relativity plus a scalar field
having exponential-like potentials \cite{R3,R4}.

Scalar-tensor theory gravity is currently of particular interest since
such theories occur as the low-energy limit in supergravity theories
from string theory and other higher-dimensional gravity theories \cite{T3}.  Lacking a full non-perturbative formulation which allows a
description of the early Universe close to the Planck time, it is
necessary to study classical cosmology prior to the GUT epoch by
utilizing the low-energy effective action induced by string theory.
To lowest order in the inverse string tension the tree-level effective
action in four-dimensions for the massless fields includes the
non-minimally coupled graviton, the scalar dilaton and an
antisymmetric rank-two tensor, hence generalizing GR (which is
presumably a valid description at late, post-GUT, epochs) by including
other massless fields; hence the massless bosonic sector of
(heterotic) string theory reduces generically to a four-dimensional
scalar-tensor theory of gravity.  As a result, BDT includes the
dilaton-graviton sector of the string effective action as a special
case $(\omega =-1)$ \cite{T3}.  String cosmology has recently
been investigated by various authors \cite{T5}, and, in particular, \cite{BCL}
presented a qualitative analysis for spatially flat, isotropic and
homogeneous cosmologies derived from the string effective action when
a cosmological constant term is included.  A discussion of how
exponential potentials arise in effective four-dimensional theories
(in the so-called conformal Einstein frame) after dimensional
reduction from higher-dimensional theories such as string theory and
M-theory is given in \cite{GZ}.

A number of authors have studied scalar field cosmological models with
an exponential potential within GR.  Homogeneous and isotropic
Friedmann-Robertson-Walker (FRW) models were studied by Halliwell \cite{R3}
using phase-plane methods (see also \cite{Olive}).  Homogeneous but
anisotropic models of Bianchi types I and III (and Kantowski-Sachs
models) have been studied by Burd and Barrow \cite{R5} in which they found
exact solutions and discussed their stability.  Lidsey \cite{R6} and
Aguirregabiria et al. \cite{R7} found exact solutions for Bianchi type I
models, and in the latter paper a qualitative analysis of these models
was also presented.  Bianchi models of types III and VI were studied
by Feinstein and Ib\'{a}\~{n}ez \cite{R8}, in which exact solutions were
found.  A qualitative analysis of Bianchi models with $k^2<2$,
including standard matter satisfying various energy conditions, was
completed by Kitada and Maeda \cite{R9}.  They found that the well-known
power-law inflationary solution is an attractor for all initially
expanding Bianchi models (except a subclass of the Bianchi type IX
models which will recollapse).

The governing differential equations in spatially homogeneous Bianchi
cosmologies containing a scalar field with an exponential potential
exhibit a symmetry \cite{S1}, and when appropriate expansion- normalized
variables are defined, the governing equations reduce to a dynamical
system, which was studied qualitatively in detail in \cite{ColeyIbanezVanDenHoogen}.  In
particular, the question of whether the spatially homogeneous models
inflate and/or isotropize, thereby determining the applicability of
the so-called cosmic no-hair conjecture in homogeneous scalar field
cosmologies with an exponential potential, was addressed. The
relevance of the exact solutions (of Bianchi types III and VI) found
by Feinstein and Ib\'{a}\~{n}ez \cite{R8}, which neither
inflate nor isotropize, was also considered.  In a follow up paper
\cite{HCI} the isotropization of the Bianchi VII$_h$ cosmological models
possessing a scalar field with an exponential potential was further
investigated; in the case $k^2>2$, it was shown that there is an open
set of initial conditions in the set of anisotropic Bianchi VII$_h$
initial data such that the corresponding cosmological models
isotropize asymptotically.  Hence, scalar field spatially homogeneous
cosmological models having an exponential potential with $k^2>2$ can
isotropize to the future. 
However, in the case of the Bianchi type IX models having an
exponential potential with $k^2>2$ the result is different in that
typically expanding Bianchi type IX models do not
isotropize to the future; the analysis of \cite{HI} indicates that if
$k^2>2$, then the model recollapses.

Recently cosmological models which contain both a perfect fluid
description of matter and a scalar field with an exponential potential
have come under heavy analysis.  One of the exact solutions found for
these models has the property that the energy density due to the
scalar field is proportional to the energy density of the perfect
fluid, hence these models have been labelled scaling cosmologies
\cite{SS,CopelandLiddleWands}.  With the discovery of these scaling
solutions, it has become imperative to study spatially homogeneous
Bianchi cosmologies containing a scalar field with an exponential
potential and an additional matter field consisting of a barotropic
perfect fluid.  The scaling solutions
studied in \cite{SS,CopelandLiddleWands},
which are spatially flat
isotropic models in which the scalar field energy density tracks that
of the perfect fluid, are of particular physical interest.  For
example, in these models a significant fraction of the current energy
density of the Universe may be contained in the scalar field whose
dynamical effects mimic cold dark matter.

In \cite{BillyardColeyVanDenHoogen} the stability of these
cosmological scaling solutions within the class of spatially
homogeneous cosmological models with a perfect fluid subject to the
equation of state $p_\gamma=(\gamma-1)\rho_\gamma$ (where $\gamma$ is
a constant satisfying $0<\gamma<2$) and a scalar field with an
exponential potential was studied.  It is known that the scaling
solutions are late-time attractors (i.e., stable) in the subclass of
flat isotropic models \cite{SS,CopelandLiddleWands}.  In
\cite{BillyardColeyVanDenHoogen} it was found that that the scaling
solutions are stable (to shear and curvature perturbations) in generic
anisotropic Bianchi models when $\gamma<2/3$.  However, when
$\gamma>2/3$, and particularly for realistic matter with $\gamma \ge
1$, the scaling solutions are unstable; essentially they are unstable
to curvature perturbations, although they are stable to shear
perturbations.  Although these solutions are unstable, since they
correspond to equilibrium points of the governing dynamical system,
the Universe model can spend an arbitrarily long time near these
scaling solutions, and hence they may still be of physical importance.

In addition to the scaling solutions described above, curvature
scaling solutions and anisotropic scaling solutions are also possible.
In \cite{HCW} homogeneous and isotropic spacetimes with non-zero
spatial curvature were studied in detail and three possible asymptotic
future attractors in an ever-expanding universe were found. In
addition to the zero-curvature power-law inflationary solution and the
zero-curvature scaling solution alluded to above, there is a solution
with negative spatial curvature where the scalar field energy density
remains proportional to the curvature, which also acts as a possible
future asymptotic attractor.  In \cite{ColeyIbanezOlasagasti}
spatially homogeneous models with a perfect fluid and a scalar field
with an exponential potential were also studied and the existence of
anisotropic scaling solutions was also discovered; the stability of
these anisotropic scaling solutions within a particular class of
Bianchi type models was discussed.

The purpose of this paper is to comprehensively study the qualitative
properties of spatially homogeneous models with a barotropic fluid and
a non-interacting scalar field with an exponential potential in the
class of Bianchi type B models (except for the exceptional case
Bianchi VI$_{-1/9}$), using the Hewitt and Wainwright formalism
\cite{WainwrightEllis,HewittWainwright}. In particular, we shall study
the generality of the scaling solutions.  The paper is organized as
follows.  In section II we define the governing equations, which are
modified from those developed in \cite{WainwrightEllis}, and discuss
the invariant sets and the existence of monotonic functions.  In section
III, we classify and list all of the equilibrium points, and their
local stability is discussed in section IV.  We give a detailed
analysis, including heteroclinic orbits, for a subset of Bianchi type
VI$_h$ models in section V.  We leave conclusions and discussion for section
VI.


\section{ The Equations}

	We shall assume that the matter content is composed of two
non-interacting components. The first component is a separately
conserved barotropic fluid with a gamma-law equation of state, i.e.,
$p=(\gamma-1)\mu$, where $\gamma$ is a constant with $0\leq\gamma\leq 2$,  
while the second is a noninteracting scalar field
$\phi$ with an exponential potential $V(\phi)=\Lambda\e^{k\phi}$, where 
$\Lambda$ and $k$ are positive constants (we use units in which $8\pi G=c=1$). By
non-interacting we mean that the energy-momentum of the two matter
components will be separately conserved. 

	The state of any Bianchi type B model with the above matter
content can be described by the evolution of the variables
\begin{equation}
  \left( H, \sigma_+, \tilde \sigma, \delta, \tilde a, n_+,\dot
	\phi,\phi\right)\in \R^8,
\end{equation} 
where the evolution of the state variables are given as equations (5.8)
and (7.8) in Wainwright and Ellis \cite{WainwrightEllis} with the addition
of the Klein-Gordon equation for the scalar field,
\begin{equation}
\ddot \phi+3H\dot\phi+kV(\phi)=0.
\end{equation}

	    By introducing dimensionless variables, the evolution equation for
$H$ decouples and the resulting reduced system has one less dimension 
\cite{WainwrightEllis}.  Defining
\cite{WainwrightEllis,ColeyIbanezVanDenHoogen} 
\begin{eqnarray}
\Sigma_+&=&\frac{\sigma_+}{H}, \quad \tilde{\Sigma} = \frac{\tilde\sigma}{H^2},
\quad \Delta= \frac{\delta}{H^2}, \quad \tilde A = \frac{\tilde a}{H^2},
\nonumber \\  N_+&=&\frac{n_+}{H},\quad \Psi=\frac{\dot \phi}{\sqrt{6}H}, \quad
\Phi=\frac{\sqrt{V(\phi)}}{\sqrt 3H},\quad \Omega=\frac{\mu}{3H^2}, \label{variable_def}
\end{eqnarray} the differential equations for the quantities 
\begin{equation}{\bf X}=
\left( \Sigma_+, \tilde{\Sigma}, \Delta, \tilde A,  N_+,\Psi,\Phi\right)\in
\R^7  \label{TheSet}\end{equation}  \ are as follows:

\begin{eqnarray}
\Sigma_+'   	&=& (q - 2)\,{\Sigma _{+}} - 2\,{\tilde N},\label{DE_1}\\
\tilde{\Sigma}'	&=& 2\,(q - 2)\,{\tilde{\Sigma}} - 4\,\Delta \,N_+ 
			- 4\,{\Sigma _{+}}\,\tilde A, \\
\Delta'    	&=& 2\,(q + {\Sigma _{+}} - 1)\,\Delta  + 2\,({\tilde{\Sigma}} 
			-{\tilde N})\,N_+, \\
\tilde{A}'   	&=& 2\,(q + 2\,\Sigma_+)\,\tilde A, \\
N_+'        	&=& (q + 2\,\Sigma_+)\,N_+ + 6\,\Delta, \\
\label{PsiDE}
\Psi'       &=& (q - 2)\,\Psi  - {\displaystyle \frac {1}{2}} \,\sqrt{6}\,k\,\Phi ^{2},\\
\label{PhiDE}
\Phi'       &=& (q+1  + {\displaystyle \frac {1}{2}} \,\sqrt{6}\,k\,\Psi 
)\,\Phi, \label{DE_7}
\end{eqnarray}
where a prime denotes differentiation with respect to the time $\tau$, where
$dt/d\tau=H$.  
The deceleration parameter $q$ is defined by $q\equiv-(1+H'/H)$, and both
$\tilde N$ (a curvature term) and $\Omega$ (a matter term) are obtained from first integrals:
 \begin{eqnarray}
  q  &=&2\,\Sigma_+^{2} + 2\,{\tilde{\Sigma}} + 
{\displaystyle \frac 
     {1}{2}}\,(3\,\gamma  - 2)\,\Omega  + 2\,\Psi ^{2} - \Phi ^{2},\\
\tilde N &=&{\displaystyle \frac {1}{3}} \,N_+^{2} - 
   {\displaystyle \frac {1}{3}} \,l\,\tilde A,\label{tilden}\\
\Omega &=& 1 - \Psi ^{2} - \Phi ^{2} - \Sigma_+^{2} - {\tilde{\Sigma}} - {\tilde N} - \tilde A .\label{constraint2}
\end{eqnarray}
The evolution of $\Omega$ is given by the auxiliary equation
\begin{equation}
\Omega'           = \Omega \,(2\,q - 3\,\gamma  + 2). \label{auxiliary}
\end{equation}
The parameter $l= 1/h$ where h is the group parameter is equivalent to
Wainwright's $\tilde h$ in \cite{WainwrightEllis}.  If $l<0$ and $\tilde A>0$ then the model 
is of Bianchi
type VI$_h$.  If $l>0$ and $\tilde A>0$ and $N_+\not= 0$ then the model is of
Bianchi type VII$_h$. If $l =0$ then the model is either Bianchi type IV or type V.  
If $\tilde A =0$ then the model is either a Bianchi type I or a Bianchi type II
model. 

There is one constraint equation that must also be satisfied:
\begin{equation}  
G({\bf X})={\tilde{\Sigma}}\,{\tilde N} - \Delta ^{2} - \tilde A\,
   \Sigma_+^{2}= 0 \label{constraint1},   
   \end{equation}
Therefore the state space is six-dimensional; the seven evolution equations
(\ref{DE_1})-(\ref{DE_7}) are subject to the constraint equation 
(\ref{constraint1}).  We shall refer to the seven-dimensional state space
(\ref{TheSet}) as the {\em extended} state space.

By definition $\tilde A$ is
non-negative, which implies from equations (\ref{constraint1}) and (\ref{tilden}) that $
\tilde{\Sigma}$ and $\tilde N$ are also non-negative.   Thus we have
\begin{equation}
\tilde A\geq0,\qquad\qquad \tilde{\Sigma}\geq0,\qquad\qquad \tilde N \geq 
0.  
\end{equation} 
In addition, from the physical constraint $\Omega\geq0$ together with equation (\ref{constraint2}), we find that the state space is compact.  Indeed, we have that
\begin{equation}
0\leq\left\{\Sigma_+^2, \tilde\Sigma, \Delta^2, \tilde A,\tilde N, \Psi^2,
\Phi \right\}\leq 1.
\end{equation}
Since both $\tilde A$ and $\tilde N$ are bounded, we have from equation (\ref{tilden}) that $N_+$ is bounded.  In equation (\ref{variable_def}) we take the ``positive square root''. 
In principle, there exists negative and positive values for $\Phi$, but
from the definition (\ref{variable_def}) a negative $\Phi$ implies a
negative $H$ and hence $H<0$ for all time; i.e., the models are
contracting.  
Since the system is invariant under $\Phi\rightarrow -\Phi$, without loss of 
generality
we shall only consider $\Phi\ge0$.


\subsection{Invariant Sets}

There are a number of important invariant sets.  Recall that the state
space is constrained by equation (\ref{constraint1}) to be a
six-dimensional surface in the seven-dimensional {\em extended} space.  
Taking the constraint equation (\ref{constraint1}) into account we calculate the dimension of each invariant set.
These invariant sets can be classified into various classes according
to Bianchi type and/or according to their matter content. Some invariant
sets (notably the Bianchi invariant sets) have lower-dimensional
invariant subsets. Equilibrium points and orbits occuring in each
Bianchi invariant set correspond to cosmological models of that
Bianchi type.  The notation used here has been adapted from
\cite{WainwrightEllis}.  Various lower-dimensional invariant sets can
be constructed by taking the intersection of any Bianchi invariant set
with the various Matter invariant sets. For example, $B(I)\cap {\cal M}$ is
a 3-dimensional invariant set describing Bianchi type I models with a
massless scalar field.
 
\begin{table}
\begin{tabular}{llcl}
Bianchi Type & Notation & Dimension  & Restrictions \\
\hline
{\it Bianchi I}  &      $B(I)$ &    4 & $\tilde{A}=\Delta=N_+=0$\\
                 &      $S(I)$ &    2 & $\tilde{A}=\Sigma_+=\tilde\Sigma=\Delta=
                                        N_+=0$ \\
\hline
{\it Bianchi II} & $B^\pm(II)$ &    5 & $\tilde{A}=0, \q N_+>0 \mbox{ or } 
                                         N_+<0$\\
                 & $S^\pm(II)$ &    4 & $\tilde{A}=0, \q \tilde\Sigma=
                                          3{\Sigma_+}^2, \q \Delta={\Sigma_+}
                                          {N_+}$ \\
\hline
{\it Bianchi IV} & $B^\pm(IV)$ &    6 & $l=0,\q \tilde{A}>0, \q N_+>0 
                                          \mbox{ or } N_+<0$\\
\hline
{\it Bianchi V}  &      $B(V)$ &    4 & $l=0,\q \tilde{A}>0, \q 
                                         \Sigma_+=\Delta=N_+=0$\\
                 &      $S(V)$ &    3 & $l=0,\q \tilde{A}>0, \q \Sigma_+=\tilde\Sigma=
                                        \Delta=N_+=0$\\
\hline
{\it Bianchi VI$_h$}&$B(VI_h)$ &    6 & $l<0, \q \tilde{A}>0$\\
                 &   $S(VI_h)$ &    4 & $l<0,\q \tilde{A}>0, \q 3\Sigma_+^2+l\tilde
					\Sigma = 0, \q 	N_+=\Delta=0$\\ 
                 &$S^\pm(III)$ &    5 & $l=-1,\q \tilde{A}>0, \q 3\Sigma_+^2-
					\tilde\Sigma = 0,\q {\Delta}=
                                        {\Sigma_+}{N_+}$\\
\hline 
{\it Bianchi VII$_h$}&$B^\pm(VII_h)$&6& $l>0,\q \tilde{A}>0, \q N_+>0 
                                        \mbox{ or } N_+<0$\\
                 & $S^\pm(VII_h)$ & 3 & $l>0,\q \tilde{A}>0,\q \Sigma_+=\tilde\Sigma=\Delta=0, 
					\q N_+^2=l\tilde{A}>0$\\
\end{tabular}
\caption{{\em Bianchi Invariant Sets. We note that $B(I)$ and
$B^\pm(II)$ are class A Bianchi invariant sets which occur in the
closure of the appropriate higher-dimensional Bianchi type B invariant
set (see Fig. 1).  In addition, if $l$ is non-negative, $N_+>0$ and
$N_+<0$ define disjoint invariant sets (indicated by a superscript
$\pm$ in the table).  Due to the discrete symmetry $\Delta\rightarrow
-\Delta$, $N_+\rightarrow -N_+$, these pairs of invariant sets are
equivalent.}\label{Table1}}
\end{table}

\begin{table}[h]
\begin{tabular}{llcl}
Matter Content  & Notation & Dimension & Restrictions \\
\hline
Scalar Field            & ${\cal S}$   & 5 & $\Omega =0$; 
	\q $\Psi\not=0,\q \Phi\not =0$ \\
Massless Scalar Field   & ${\cal M}$   & 4 & $\Omega =0$; 
	\q $\Psi\not =0,\q \Phi =0$ \\
Vacuum                  & ${\cal V}$     & 3 & $\Omega =0$; 
	\q $\Psi=0,\q \Phi =0$ \\
\hline 
Perfect Fluid + Scalar Field        & ${\cal FS}$ & 6 & $\Omega\neq 0$;
	\q $\Psi\not =0,\q \Phi \not =0$  \\ 
Perfect Fluid + Massless Scalar Field     & ${\cal FM}$ & 5 & $\Omega\neq 0$;
	\q $\Psi\not=0,\q \Phi =0$  \\
Perfect Fluid                       & ${\cal F}$     & 4 & $\Omega\neq 0$;
	\q $\Psi=0,\q \Phi =0$ 
\end{tabular}
\caption{{\em Matter Invariant Sets.}\label{Table2}}
\end{table} 

An analysis of the dynamics in the 
invariant 
sets ${\cal V}$ and ${\cal F}$ has been presented by Wainwright and
 Hewitt \cite{HewittWainwright}.  Equilibrium points and orbits in the
 invariant set ${\cal M}$ correspond to models with a massless scalar
 field; i.e., scalar field models with zero potential.  These models
 are equivalent to models with a stiff perfect fluid 
(i.e. $\gamma=2$)
 equation of state; see \cite{HewittWainwright}.  Equilibrium points
 and orbits in the invariant set ${\cal FM}$ can be interpreted as
 representing a two--perfect--fluid model with $\gamma_2=2$
 \cite{ColeyWainwright}. A partial analysis of the isotropic
 equilibrium points in the 
invariant
set ${\cal S}$ was completed by van den
 Hoogen et al. \cite{VanDenHoogenColeyIbanez}.  We note that the
 so-called scaling solutions
 \cite{CopelandLiddleWands,Wetterich,Wands} are in the invariant set
 ${\cal FS}$.

The isotropic and spatially homogeneous models are found in the invariant
sets
 $S^\pm(VII_h) \cup S(I)$ if $l\not =0$, and $S(V) \cup S(I)$ if $l=0$. In
particular the zero curvature isotropic models are found in the two
dimensional set $S(I)$, while the negative curvature models are found in
the three-dimensional sets $S^\pm(VII_h)$ or $S(V)$ depending upon the value of $l$.  See van den Hoogen et al.
for a comprehensive analysis of the isotropic scaling models \cite{HCW}.

We note that in the invariant set $B(I)$ there exists the invariant set 
$\tilde{\Sigma}+\Sigma_+^2+\Psi^2<1, 
\Delta=\tilde{A}=N_+=\Phi=0$, 
which may be directly integrated to yield
\begin{equation}
\tilde{\Sigma}+\Sigma_+^2+\Psi^2 = \left[1+\zeta e^{3(2-\gamma)\tau} 
\right]^{-1}, \q \zeta=\mbox{constant},
\end{equation}
where $\tau$ is the time parameter.  This solution asymptotes into the
past towards the paraboloid ${\cal K}$ (section III.B.1), and
asymptotes to the future towards the point
P$(I)$.  This
solution
belongs to the matter invariant set ${\cal FM}$, asymptoting into the
past towards the set ${\cal M}$.


\subsection{Monotone Functions}

The existence of strictly monotone functions, $W({\bf X}):\R^n\to\R$,
on any invariant set, $S$, proves the non-existence of periodic or
recurrent orbits in $S$ and 
can be used to provide 
 information about the global
behaviour of the dynamical system in $S$. (See Theorem 4.12 in
\cite{WainwrightEllis} for details.)

\begin{table}[h]
\begin{tabular}{lll}
Function: $W_i({\bf X})$  & Derivative: $W_i'({\bf X})$ & Region of Monotonicity  \\
\hline
	& & Monotonically approaches zero \\
$\displaystyle W_1\equiv \left(1+\Sigma_+\right)^2-\tilde A$ &
	$\displaystyle W_1' = -2\left(2-q\right) 	
	W_1+3\left(1+\Sigma_+\right)\left(2\Phi^2+\left(2-\gamma\right)\Omega\right)$ &
	in the invariant set $\calm \cup \calv $.  \\
	\\
\hline
	& & Monotonically decreasing to zero \\
$\displaystyle W_2\equiv \frac{1-\Omega -\Phi^2-\Psi^2}{\Omega}$ & 
	$\displaystyle W_2' = -W_2(2-3\gamma) - \frac{1}{\Omega}\left(\Sigma_+^2+\tilde\Sigma\right)$ & 
	in the set $({\cal FS} \cup {\cal FM} \cup {\cal F})\backslash S(I)$ \\
	& & when $0\leq \gamma\leq 2/3$\\
\hline
	& & Monotonically decreasing to zero \\
$\displaystyle W_3\equiv \tilde{\Sigma}$ &
	$\displaystyle W_3' = -2\left(2-q\right) W_3-4(\Delta N_+ +\Sigma_+\tilde{A})$ &
 	in the invariant sets $B(I)\backslash S(I)$ and $B(V)\backslash S(V)$. \\
 	\\
\hline
	& & Monotonically approaches zero \\
$\displaystyle W_4\equiv \frac{\tilde{A}^2}{N_+} $ &
	$\displaystyle W_4' = 3W_4\left(q+2\frac{\Sigma_+N_+-\Delta}{N_+}\right)$ &
 	in the invariant set $S^{\pm}(III)\backslash({\cal S}\cup{\cal FS})$,\\
 	& & when $\gamma>2/3$.
\end{tabular}
\caption{{\em Functions, their derivatives and the sets in which they are monotonic.}\label{Table3}}
\end{table}

Hewitt and Wainwright found a number of monotone functions in the
invariant sets of dimension less than four in the perfect fluid case
(i.e., in lower-dimensional subsets of the perfect fluid invariant
set) and these are summarized in an Appendix in Hewitt and Wainwright
\cite{WainwrightEllis,HewittWainwright}.  However, they were not able
to find a monotonic function in the full perfect fluid invariant set
for $2/3<\gamma<2$.


\subsection{The Constraint Surface}

The constraint equation $G({\bf X})=0$ and the Implicit Function
Theorem can generally be used to eliminate one of the variables at any
point in the {\em extended} state-space provided the constraint
equation is not singular there, i.e., $\grad(G({\bf X}))\not = {\bf
0}$.  The constraint surface is singular for all points in the
invariant sets $S(I)$, $B(V)$ and $S(VII_h)$ and therefore cannot be
used to eliminate one of the variables (and hence reduce the dimension
of the dynamical system to six).

Therefore, we cannot determine the local stability of equilibrium
points in the sets $S(I)$, $B(V)$ or $S(VII_h)$ within the
six-dimensional state-space, and hence we are required to determine
the local stability of these equilibrium points in the {\em extended}
space, due to the singular nature of the constraint surface. This
leads to further complications because of the limited use of the
Stable Manifold Theorem.  If these equilibrium points are stable in
the {\em extended} state space, then they are stable in the
six-dimensional constrained surface.  
However, if these equilibrium
points are saddles in the {\em extended} state-space, then one
cannot easily determine the dimension of the stable manifold within the
constraint surface.
   

\section{Classification of the Equilibrium Points} 

Let us analyse the evolution equations for the matter variables, namely
equations (\ref{PsiDE}) and (\ref{PhiDE}) and the auxiliary equation 
(\ref{auxiliary}).  From equation (\ref{auxiliary}) we find that at the 
equilibrium points either
\begin{equation}
(A) \quad \Omega  =  0, \label{caseA}
\end{equation} 
or 
\begin{equation}
(B) \quad q  =  \frac{3}{2}\gamma-1. \label{caseB}
\end{equation}
In the scalar field case $(A)$ there is no perfect fluid present.
This is the scalar field invariant set ${\cal S}$.  The equilibrium
points and their stability will be studied in subsection III.A.  These
models include the massless scalar field case in which $\Phi=0$
($V=0$), but not the vacuum case $\Phi=\Psi=0$ which will be dealt with
as a subcase of the perfect fluid case (see below).  The equilibrium
points of case $(A)$ include the isotropic Bianchi VII$_h$ models
studied in \cite{VanDenHoogenColeyIbanez}.

If, on the other hand, equation (\ref{caseB}) is satisfied, assuming that
$\gamma<2$ so that $q\neq 2$, from equations (\ref{PsiDE}) and (\ref{PhiDE}) 
we have that
\begin{equation}
(B1) \quad \Psi=0, \Phi=0 \label{caseB1}\end{equation}
or \begin{equation}
(B2) \quad \Psi=\frac{-\sqrt 3 \gamma}{\sqrt 2 k}, \quad
     \Phi^2=\frac{3\gamma(2-\gamma)}{2k^2} \label{caseB2}.
\end{equation}
In case $(B1)$, in which both equations (\ref{caseA}) and (\ref{caseB}) are 
valid, there is no scalar field present.  The perfect fluid subcase, which
was studied by Hewitt and Wainwright \cite{HewittWainwright}, will be dealt
with in subsection III.B.  Note that from equation (\ref{PhiDE}) $\Phi=0$
is an invariant set, denoted ${\cal M}$.

The final case $(B2)$, in which equation (\ref{caseB}) is valid and neither
the scalar field nor the perfect fluid is absent, corresponds to the scaling
solutions when $\gamma>0$.  If we define
\begin{equation}
\mu_\phi \equiv \frac{1}{2}\dot\phi^2 +V(\phi), \quad 
	  p_\phi \equiv \frac{1}{2} \dot\phi^2-V(\phi), \label{phiFluid}
\end{equation}
then from equation (\ref{caseB2}) we find that
\begin{equation}
\gamma_\phi \equiv \frac{\mu_\phi+p_\phi}{p_\phi} = 
	\frac{2\Psi^2}{\Psi^2+\Phi^2} = \gamma,
\end{equation}
so that the scalar field ``inherits'' the equation of state of the fluid.  It
can be shown that there are exactly three equilibrium points corresponding to
scaling solutions; the flat isotropic scaling solution described in 
\cite{CopelandLiddleWands}, and whose stability was discussed within Bianchi 
type 
VII$_h$ models in \cite{BillyardColeyVanDenHoogen}, and two anisotropic scaling 
solutions \cite{ColeyIbanezOlasagasti}.  This will be further discussed in 
subsection III.C.

Hereafter, we shall assume that $0<\gamma<2$.  The value $\gamma=0$
corresponds to a cosmological constant and the model can be analyzed as
a scalar field model with the potential $V=V_0+\Lambda e^{k\phi}$
\cite{BillyardColeyIbanez}.  The value $\gamma=2$, corresponding to the
stiff fluid case, is a bifurcation value and will not be considered further. 


\subsection{Scalar Field Case} 

There are 
seven equilibrium points and one equilibrium set 
in the scalar field
invariant set ${\cal S}$ in which $\Omega=0$.  
The first three equilibrium points were given in
\cite{ColeyIbanezVanDenHoogen}
(wherein matter terms were not included):
 they represent isotropic
models ($\Sigma_+=\tilde \Sigma=\tilde N=\Delta=0$):

\

\noindent{\bf 1)} P$_{\cal S}(I)$: $\Sigma_+=\tilde{\Sigma}=\Delta=\tilde{A}=N_+=0, \Psi=-k/\sqrt{6}, \Phi=\sqrt{1-k^2/6}$
    
This equilibrium point, for which $q=-1+k^2/2$ and which exists only
for $k^2\leq 6$, is in the Bianchi I invariant set $B(I)$.  This
point represents a flat FRW power-law inflationary model
\cite{R3,ColeyIbanezVanDenHoogen}.  The corresponding eigenvalues in
the extended state space are (throughout this paper, we shall not
explicitly display the corresponding eigenvectors):
\begin{equation}
-\half (6-k^2), \q-\half (6-k^2), \q  -(6-k^2), \q -(4-k^2), \q -(2-k^2),
\q -\half(2-k^2), \q k^2-3\gamma.
\end{equation}

\

\noindent{\bf 2)} P$^{\pm}_{\cal S}(VII_h)$: $\Sigma_+=\tilde{\Sigma}=\Delta=0, \tilde{A}=\frac{(k^2-2)}{k^2},
     N_+=\pm\frac{\sqrt{l(k^2-2)}}{k},\Psi=-\frac{\sqrt{2}}{\sqrt{3}k}, 
     \Phi=\frac{2}{\sqrt{3}k}$

These two equilibrium points (the indices ``$\pm$'' correspond to the $\pm$
values for 
$N_+$), which occur in the Bianchi VII$_h$ invariant set $S(VII_h)$
(since $\tilde{A}\geq0$, then $k^2\geq 2$ and therefore $l>0$), have $q=0$.  These
equilibrium points
represent an open FRW model \cite{VanDenHoogenColeyIbanez}.  The
corresponding eigenvalues in the extended state space are:
\begin{equation}
2-3\gamma, \q -1\pm\frac{\sqrt{3}i}{k}\sqrt{k^2-8/3}, -2\pm\frac{\sqrt 2}{k}
\sqrt{k^2-4(k^2-2)l\pm\sqrt{\left[k^2-4(k^2-2)l\right]^2+16l\left(k^2-2\right)^2+k^4}}.
\end{equation}

\

{\bf 2a)} P$_{\cal S}(V)$: $\Sigma_+=\tilde{\Sigma}=\Delta=0, 
\tilde{A}=\frac{(k^2-2)}{k^2},
     N_+=0,\Psi=-\frac{\sqrt{2}}{\sqrt{3}k}, 
     \Phi=\frac{2}{\sqrt{3}k}$

This case corresponds to points 2) for $l=0$ and belongs to the set $S(V)$. 
The corresponding eigenvalues in the extended state space are:
\begin{equation}
2-3\gamma, \q -1\pm\frac{\sqrt{3}i}{k}\sqrt{k^2-8/3}, \q -2,\q -2, \q
0, \q -4.
\end{equation}

\

%

\noindent{\bf 3)} P$^{\pm}_{\cal S}(II)$: $\Sigma_+=-\frac{k^2-2}{k^2+16}, 
   \tilde{\Sigma}=3\Sigma_+^2, \Delta=\Sigma_+N_+,\tilde{A}=0, 
   N_+=\pm 3\frac{\sqrt{-(k^2-2)(k^2-8)}}{k^2+16},
    \Psi=-\frac{3\sqrt6 k}{k^2+16},\Phi=6\frac{\sqrt{8-k^2}}{k^2+16}$

These two equilibrium points, for which $q=8(k^2-2)/(k^2+16)>0$, exist
only for $2\leq k^2 \leq 8$.  These two points represent Bianchi type
II models analogous to those found in \cite{HewittWainwright}.  The
corresponding eigenvalues are:
\begin{equation}
12\frac{k^2-2}{k^2+16}, \q 6\frac{k^2-8}{k^2+16},
   \q 6\frac{k^2-8}{k^2+16}, \q 3\frac{(k^2-8)\pm \sqrt{(13k^2-32)(k^2-8)}}
   {k^2+16}, \q -3\gamma+18\frac{k^2}{k^2+16}.
\end{equation}

\

\noindent{\bf 4)} P$_{\cal S}(VI_h)$: $\Sigma_+=\frac{-l(k^2-2)}{n}, 
    \tilde{\Sigma}=-3\Sigma_+^2/l,\Delta=0,\tilde{A}=\frac{9(k^2-2l)(k^2-2)}
     {n^2}, N_+=0, \Psi=\frac{\sqrt 6k(1-l)}{n},
    \Phi=\frac{2\sqrt 3 \sqrt{(k^2-2l)(1-l)}}{n},$
where $n\equiv k^2(l-3)+4l$.  Since $\tilde{\Sigma}>0$, we have that
$l<0$ and hence 
this equilibrium point occurs
in the Bianchi
VI$_h$ invariant sets.  The deceleration parameter is given by
$q=2l(k^2-2)/[k^2(l-3)+4l]\geq0$, where $k^2\geq2$, and 
this point corresponds to a Collins Bianchi type VI$_h$ solution \cite{Collins}.
The corresponding eigenvalues are:
\begin{eqnarray} \nonumber
&& 6\frac{k^2-2l}{[k^2(l-3)+4l]}, \q
  -3\gamma-6\frac{k^2(1-l)}{[k^2(l-3)+4l]}, \\
&& 3\frac{(k^2-2l)\pm\sqrt{(k^2-2l)^2+8l(1-l)(k^2-2)}}{[k^2(l-3)+4l]}, \q
   3\frac{(k^2-2l)\pm\sqrt{(k^2-2l)[(k^2-2l)-4(1-l)(k^2-2)]}}{[k^2(l-3)+4l]}.
\end{eqnarray}

\

Let us next consider {\em the Massless Scalar Field Invariant Set
${\cal M}$}: there is 
one equilibrium set which generalizes the
work in \cite{HewittWainwright} to include scalar fields:

\

\noindent{\bf 5)} ${\cal K}_{\cal M}$: $\tilde{\Sigma}+\Sigma_+^2+\Psi^2=1, 
\Delta=\tilde{A}=N_+=\Phi=0, \Psi\neq 0$

This paraboloid, for which $q=2$, generalizes the parabola ${\cal K}$
in \cite{HewittWainwright} defined by $\tilde{\Sigma}+\Sigma_+^2=1$
to include a massless scalar field, and represents Jacobs' Bianchi type I
non-vacuum solutions \cite{Collins}.  However, the eigenvalues are
considerably different from those found in \cite{HewittWainwright},
and so we list them all here 
(the variables which define the subspaces in which the corresponding eigendirections reside are included below in curly braces):
\begin{equation}
{2[(1+\Sigma_+)\pm\sqrt{3\tilde\Sigma}], \atop \{\Delta, 
N_+\}} \q
{0, \atop \{\Sigma_+, \tilde{\Sigma}\}} \q {0, \atop \{\Sigma_+, 
\tilde{\Sigma},\Psi\}} \q
{3(2-\gamma), \atop \{\Sigma_+,\tilde{\Sigma},\Psi\}} \q
{4(1+\Sigma_+), \atop \{\Sigma_+,\tilde{\Sigma},\tilde{A},\Psi\}} \q
{\frac{\sqrt 6}{2}\left( \sqrt{6} +
k\Psi\right). \atop \{\tilde{\Sigma},\Phi\}} 
\end{equation}


\subsection{Perfect Fluid Case, $\Psi=\Phi=0$} 

As mentioned earlier, the perfect fluid invariant set ${\cal F}$ in
which $\Psi=\Phi=0$ was studied by Hewitt and Wainwright
\cite{HewittWainwright}; hence this subsection generalizes their
results by including a scalar field with an exponential potential.  We
shall use their notation to label the equilibrium points/sets.  There
are five such invariant points/sets.  In all of these cases the extra
two eigenvalues associated with $\Psi$ and $\Phi$ are (respectively)
\begin{equation} -\frac{3}{2}(2-\gamma)<0, \q \frac{3}{2}\gamma>0. 
\end{equation}

\

\noindent{\bf 1)} P(I): 
$\Sigma_+=\tilde{\Sigma}=\Delta=\tilde{A}=N_+=\Psi=\Phi=0$

This equilibrium point, for which $\Omega=1$, is a saddle for
$2/3<\gamma<2$ in ${\cal F}$ \cite{HewittWainwright} (and is a sink
for $0\le\gamma<2/3$), which corresponds to a flat FRW model.

\

\noindent{\bf 2)} P$^\pm$(II): $\Sigma_+=-\frac{1}{16}(3\gamma-2), 
  \tilde{\Sigma}=3\Sigma_+^2, \Delta=\Sigma_+N_+, \tilde{A}=0, 
N_+=\pm\frac{3}{8}
   \sqrt{(3\gamma-2)(2-\gamma)}, \Psi=\Phi=0$

This equilibrium point, for which $\Omega=\frac{3}{16}(6-\gamma)$, is a saddle
in the perfect fluid invariant set \cite{HewittWainwright}. 

\

\noindent{\bf 3)} P(VI$_h$): $\Sigma_+=-\frac{1}{4}(3\gamma-2), \tilde{\Sigma}=-
  3\Sigma_+^2/l, \Delta=0, 
\tilde{A}=-\frac{9}{16l}(3\gamma-2)(2-\gamma),
 N_+= \Psi=\Phi=0$

Since $\tilde{\Sigma}>0$ and $\tilde{A}>0$, this equilibrium point
occurs in the Bianchi VI$_h$ invariant set and corresponds to the
Collins solution \cite{Collins}, where $\Omega=\frac{3}{4}(2-\gamma)
+\frac{3}{4l}(3\gamma-2)$ (and therefore $2/3\leq\gamma\leq 2(-l-1)/(3-l)$
and so $l\leq -1$). In \cite{HewittWainwright} this was a sink in
${\cal F}$, but is a saddle in the extended state space due to the
fact that the two new eigenvalues have values of different sign.

\

There are also two equilibrium sets, which generalize the work in 
\cite{HewittWainwright} to include scalar fields:

\

\noindent{\bf 4)} ${\cal L}^\pm_l$: $\tilde{\Sigma}=-\Sigma_+(1+\Sigma_+), 
\Delta=0,
     \tilde{A}=(1+\Sigma_+)^2,
     N_+=\pm\sqrt{(1+\Sigma_+)[l(1+\Sigma_+)-3\Sigma_+]}, \Psi=\Phi=0$

For this set $\Omega=0$. The local sinks in this set 
occur when \cite{HewittWainwright} 

(a) $l<0$ (Bianchi type VI$_h$) for $-\frac{1}{4}(3\gamma-2)<\Sigma_+ 
<l/(3-l)$ and $l>-(3\gamma-2)/(2-\gamma)<0$,

(b) $l=0$ (Bianchi type IV) for $-\frac{1}{4}(3\gamma-2)<\Sigma_+ < 0$,

(c) $l>0$ (Bianchi type VII$_h$) for $-\frac{1}{4}(3\gamma-2)<\Sigma_+ < 0$.
  
\noindent The additional two eigenvalues for the full system are:
\begin{equation}
1-2\Sigma_+,\q -2(1+\Sigma_+).
\end{equation}

Finally, let us consider {\em the Massless Scalar Field Invariant Set ${\cal FM}$}:

\noindent{\bf 5)} ${\cal K}$: $\tilde{\Sigma}+\Sigma_+^2=1, 
\Delta=\tilde{A}=N_+=\Phi=\Psi=0$

This parabola, for which $q=2$, is the special case of ${\cal K}_{\cal
M}$ for which $\Psi=0$ and corresponds to the parabola ${\cal K}$ in
\cite{HewittWainwright}.  However, the eigenvalues are considerably
different from those found in \cite{HewittWainwright} and so we list
them all here 
(the variables define the subspaces in which the corresponding eigendirections reside are included below in curly braces):
\begin{equation}
{2[(1+\Sigma_+)\pm\sqrt{3\tilde{\Sigma}}], \atop \{\Delta, 
N_+\}} \q
{0, \atop \{\Sigma_+, \tilde{\Sigma}\}} \q {0, \atop \{\Psi\}} \q
{3(2-\gamma), \atop \{\Sigma_+,\tilde{\Sigma}\}} \q
{4(1+\Sigma_+), \atop \{\Sigma_+,\tilde{\Sigma},\tilde{A}\}} \q
{3. \atop \{\Phi\}} 
\end{equation}

We include here the equilibrium points/sets and corresponding
eigenvalues as listed in \cite{HewittWainwright}.
\begin{table}[ht]
\begin{tabular}{lcl}
Eqm. point/set & Eigenvalues & Comment \\
\hline \hline
$P(I)$ & $-\frac{3}{2}(2-\gamma) \q -3(2-\gamma) \q (3\gamma -4) \q 
\frac{1}{2}(3\gamma-2) \q \frac{1}{2}(3\gamma-2)$ &  \\
\hline
$P^\pm(II)$ & $\frac{3}{4}(3\gamma-2) \q -\frac{3}{2}(2-\gamma) \q -\frac{3}{4}(2-\gamma)\left\{1\pm\sqrt{1-\frac{(3\gamma-2)(6-\gamma)}{2(2-\gamma)}}\right\}$ & Constraint eqn. used to eliminate $\tilde\Sigma$ \\
\hline
$P(VI_h) ^\dagger$ & $-\frac{3}{4}(2-\gamma)(1\pm\sqrt{1-r^2}) \q 
-\frac{3}{4}(2-\gamma) (1\pm \sqrt{1-q^2})$ & Constraint eqn. used to eliminate $\tilde\Sigma$\\
\hline
${\cal K}$ & $0 \q 2(1+\Sigma_+) \q 2(2-\gamma) \q 2\left[1+\Sigma_+ \pm
\sqrt{3(1-\Sigma_+^2)} \right]$ & 1-D invariant set \\
\hline
${\cal D}$ & $ 0 \q 0 \q 2
\left[1+\Sigma_+\pm\sqrt{3\tilde\Sigma}\right] \q
2\left(1+\Sigma_+\right) $ & 2-D invariant set, $\gamma=2$\\
\hline
${\cal L}_l$ & $0 \q -4\Sigma_+-(3\gamma-2) \q -2\left[ (1+\Sigma_+) \pm
2iN_+ \right] $ & Constraint eqn. used to eliminate $\tilde\Sigma$ \\
\hline
${\cal F}_l$ & $-2 \q 4 \q -2 \q 0 \q 0$ & $l\geq 0$, non-hyperbolic, $\gamma=2/3$ 
\end{tabular}
\caption{{\em Equilibrium sets found by Hewitt and Wainwright [26],
and the corresponding eigenvalues in the extended space.  In the table
$r^2\equiv 2(3\gamma-2)(1-l_c/l)$, $q^2\equiv 2r^2/(2-\gamma)$ and
$l_c\equiv -(3\gamma-2)/(2-\gamma)$.}}
\end{table}


\subsection{Scaling Solutions} 
Defining 
\begin{equation}
\Psi_S\equiv -\sqrt{\frac{3}{2}} \frac{\gamma}{k}, \quad 
    \Phi^2_S\equiv \frac{3\gamma(2-\gamma)}{2k^2},
\end{equation}
and recalling that $0<\gamma<2$, there are three equilibrium points
corresponding to scaling solutions.  Because the scalar field mimics
the perfect fluid with the exact same equation of state
($\gamma_\phi=\gamma$) at these equilibrium points, one may combine
these two ``fluids'', via $p_{tot}=p_\phi+p$,
$\mu_{tot}=\mu_\phi+\mu$, $p_{tot}=(\gamma-1)\mu_{tot}$; therefore,
all of these equilibrium points will correspond to exact perfect fluid
models analogous to the equilibrium points found in
\cite{HewittWainwright}.

\

The flat isotropic FRW scaling solution \cite{Wetterich,Wands}:

\noindent{\bf 1)} ${\cal F}_S(I)$: $\Sigma_+=\tilde{\Sigma}=\Delta=A=N_+=0,
\Psi=\Psi_S, \Phi=\Phi_S$

The eigenvalues for these points in the extended space, for which 
$\Omega=1-3\gamma/k^2$ (and therefore $k^2\geq 3\gamma$) are:
\begin{equation}
-\frac{3}{2}(2-\gamma), \q -3(2-\gamma), \q 3\gamma-4, \q 3\gamma-2, \q
\frac{1}{2}(3\gamma-2), -\frac{3}{4}(2-\gamma)\pm
\frac{3}{4}\sqrt{(2-\gamma)(2-9\gamma +24\gamma/k^2)} 
\end{equation}

There are two anisotropic scaling solutions:

\

\noindent{\bf 2)} ${\cal A}_S(II)$: $\Sigma_+=-\frac{1}{16}{(3\gamma-2 )}, 
\tilde\Sigma=3\Sigma_+^2,
\Delta=\Sigma_+N_+, \tilde{A}=0, N_+=\pm\frac{3}{8}\sqrt{(3\gamma-2)
(2-\gamma)},\Psi=\Psi_S, \Phi=\Phi_S$

The eigenvalues for these points, for which
$\Omega=\frac{3}{16}(6-\gamma)-3\gamma/k^2$ (and therefore
$k^2\geq16\gamma/[6-\gamma]$), are:
\begin{eqnarray}\nonumber &&
\frac{3}{4}\left(3\gamma-2\right), \q -\frac{3}{2}(2-\gamma),\\
	&&-\frac{3}{4}\left[(2-\gamma)
	\pm\sqrt{ \left(2-\gamma\right)^2 -\frac{3}{4}\left(2-\gamma\right) 
          \left\{
	  2\left(3\gamma-2\right) +\frac{\gamma(6-\gamma)}{k^2}\left(
	  k^2 -\frac{16\gamma}{6-\gamma}\right) \pm \sqrt{E_1} \right\}}\right],
\end{eqnarray}
where $E_1\equiv \left[
	  2\left(3\gamma-2\right) -\frac{\gamma(6-\gamma)}{k^2}\left(
	  k^2 -\frac{16\gamma}{6-\gamma}\right)\right]^2 +\frac{8}{9}\left(3\gamma-2\right)\frac{\gamma(6-\gamma)}{k^2}\left(
	  k^2 -\frac{16\gamma}{6-\gamma}\right) .$

\

\noindent{\bf 3)}${\cal A}_S(VI_h)$:
$\Sigma_+=-\frac{1}{4}\left(3\gamma- 2\right),
 \tilde\Sigma=-3\Sigma_+^2/l,\Delta=0,\tilde{A}=-\frac{9}{16l}(2-\gamma)
\left(3\gamma-2\right), N_+=0,\Psi=\Psi_S,\Phi=\Phi_S$

These points occur in the Bianchi VI$_h$ invariant set ($l<0$ since
$\tilde{\Sigma}>0$) for which
$\Omega=\frac{3}{4}(2-\gamma)+\frac{3}{4l}(3\gamma-2) -3\gamma/k^2$
(and therefore $-l^{-1}\leq (2-\gamma)/(3\gamma-2)$ and $k^2\geq 4\gamma/[(2-\gamma)+(3\gamma-2)/l]$ ) and
correspond to the Collins Bianchi VI$_h$ perfect fluid solutions
\cite{Collins}.  The eigenvalues for these equilibrium points are:
\begin{eqnarray}
\nonumber
&&-\frac{3}{4}\left[(2-\gamma)\pm
\sqrt{(2-\gamma)^2-4(3\gamma-2)^2\left(\frac{2-\gamma}{3\gamma-2}+\frac{1}{l}   \right)}\right], 
\\
\label{bugger}
&& -\frac{3}{4}\left[(2-\gamma)\pm
	\sqrt{(2-\gamma)^2-(2-\gamma)\left[
	4\gamma\left(1-\frac{3\gamma}{k^2}\right)+\left(3\gamma-2\right)
	\left(\frac{2-\gamma}{3\gamma-2}+\frac{1}{l}\right) \pm \sqrt{E_2} 
        \right]}\right],
\end{eqnarray}
where $E_2\equiv
	\left[4\gamma\left(1-\frac{3\gamma}{k^2}\right)-\left(3\gamma-2\right)
	\left(\frac{2-\gamma}{3\gamma-2}+\frac{1}{l}\right)\right]^2
	-128\frac{\gamma^2}{k^2}$.

\section{Stability of the Equilibrium Points and Some Global Results}

The stability of the equilibrium points listed in the previous section
can be easily determined from the eigenvalues displayed.  Often the
stability can be determined by the eigenvalues in the extended state
space, otherwise the constraint must be
utilized to determine the stability in the six-dimensional state space
(i.e., within the constraint surface).  In the cases in which this is not
possible, we must analyse the eigenvalues in the extended seven-dimensional 
state-space, and the conclusions that can
be drawn are consequently limited.  Employing local stability results and utilizing
the monotone functions found in Table \ref{Table3}, we
are able to prove some global results.  In the absence of monotone
functions, and in the same spirit as Refs. \cite{WainwrightEllis} and
\cite{HewittWainwright}, we conjecture plausible results which are
consistent with the local results and the dynamical behaviour on the
boundaries and which are substantiated by numerical experiments.

\subsection{The Case $\Omega=0$}

If $\Omega=0$ and $\Phi=0$, then the function $W_1$ in Table
\ref{Table3} monotonically approaches zero.  The existence of the monotone function $W_1$ implies that the
global behaviour of models in the set ${\cal M}\cup{\cal V}$ can be
determined by the local behaviour of the equilibrium points in ${\cal
M}\cup{\cal V}$.  Consequently, a portion of the equilibrium sets ${\cal
K}$ and ${\cal K}_{\cal M}$ (corresponding to local sources) represent
the past asymptotic states while the future asymptotic state is
represented by ${\cal L}_l$, or in the case of Bianchi types I and
II, by a point on ${\cal K}$.

Therefore, all vacuum models and all massless scalar field models are
asymptotic to the past to a Kasner state and are asymptotic to the
future either to a plane wave solution (Bianchi types IV, VI$_h$ and
VII$_h$), or to a Kasner state (Bianchi types I and II), or to a Milne
state (Bianchi type V).

If $\Omega=0$ and $\Phi\neq0$, then the models only contain a scalar
field.  It was proven in \cite{R9} that all Bianchi models evolve to a
power-law inflationary state (represented by P$_{\cal S}(I)$) when
$k^2<2$.  If $k^2>2$, then it was shown in
\cite{VanDenHoogenColeyIbanez} that a subset of Bianchi models of
types V and VII$_h$ evolve towards negatively curved isotropic models
represented by points P$_{\cal S}(V)$ and P$^\pm_{\cal S}(VII_h)$.  In
\cite{HCI} it was shown that when $k^2>2$ the future state of a subset
of Bianchi type VI$_h$ solutions is represented by the point P$_{\cal
S}(VI_h)$.  It can be seen here that the future state of a subset of
Bianchi type II models is represented by the point P$^\pm_{\cal
S}(II)$.  

Therefore, all scalar field models with $\Omega=0$ evolve to a
power-law inflationary state if $k^2<2$.  If $k^2>2$, then the future
asymptotic state for all Bianchi type IV, V and VII$_h$ models is
conjectured to be a negatively--curved, isotropic model and the future
asymptotic state for all Bianchi type VI$_h$ models is conjectured to
be the Feinstein-Ib\'{a}\~{n}ez anisotropic scalar field model
\cite{R8}.  If $2<k^2<8$, then the future asymptotic state for all
Bianchi type II models is the anisotropic Bianchi type II scalar field
model, and if $k^2\geq8$ then the future asymptotic state is that of a
Kasner model.  If $2<k^2<6$, then the Bianchi type I models approach a
non-inflationary, isotropic (i.e., the point P$_{\cal S}(I))$; if
$k^2\geq 6$, then they evolve to a Kasner state in the future.

\subsection{The Case $\Omega\neq0$, $0\leq\gamma\leq2/3$}

If $\Omega\neq0$ and $0\leq\gamma\leq2/3$ then the function $W_2$ in
Table \ref{Table3} is monotonically decreasing to zero.  Therefore, we
conclude that the omega-limit set of all non-exceptional orbits (i.e.,
those orbits excluding equilibrium points, heteroclinic orbits, etc.)
of the dynamical system (\ref{DE_1})-(\ref{DE_7}) is a subset of
$S(I)$.  This implies that all non-exceptional models with
$\Omega\neq0$ evolve towards the zero--curvature
spatially--homogeneous and isotropic models in $S(I)$ and hence
isotropize to the future.  In \cite{HCW}, it was shown that the
zero--curvature spatially--homogeneous and isotropic models evolve
towards the power-law inflationary model, represented by the point
P$_{\cal S}(I)$ when $k^2<3\gamma$ or towards the isotropic scaling
solution, represented by the point ${\cal F}_{\cal S}(I)$, when
$k^2>3\gamma$.  Using $W_1$, we also conclude that the past asymptotic
state(s) of all non-exceptional models (including models in $S(I)$) is
characterized by $\Omega=0$.  In other words, matter is dynamically
{\em unimportant} as these models evolve to the past.  It was shown in
\cite{HCW} that all models evolve in the past to some portion of
${\cal K}$ or ${\cal K}_{\cal M}$ (the Kasner models) which are local
sources.


\subsection{The case $\Omega\neq0$, $\frac{2}{3}<\gamma<2$}

The following table lists the local sinks for $\frac{2}{3}<\gamma<2$.
\pagebreak
\begin{table}[ht]
\begin{tabular}{lccc}
Sink   & Bianchi type & k & Other constraints \\ \hline\hline
P$_{\cal S}(I) $              & I       & $k^2\leq 2$ & \\
P$_{\cal S}(VII_h)^\dagger$   & I       & $k^2=2$ & \\ \hline
P$_{\cal S}(VI_{-1}) $    & III     & $k^2\geq 2$ & $\gamma>k^2/(k^2+1),\q l=-1$  \\
${\cal L}^\pm_k(VI_{-1}) $    & III     & all &  $\gamma>1,\q \Sigma_+=-1/4$\\ \hline
P$_{\cal S}(V)$               & V       & $k^2\geq2$ & \\ \hline
P$_{\cal S}(VI_h) $       & VI$_h$  & $k^2\geq 2$ & $\ \ \gamma>2k^2(1-l)/[k^2(l-3)+4l]$ \\
${\cal L}^\pm_k(VI_h) $       & VI$_h$  & all &$\gamma>4/3,\q \Sigma_+<-1/2$\\ 
${\cal A_S}(VI_h)$	      & VI$_h$  & $k^2 \geq \frac{ 4\gamma} 
	{[(2-\gamma)+(3\gamma-2)/l]}$ & $l \leq \frac{-(3\gamma-2)}{2-\gamma}$
 \\ \hline
P$^\pm_{\cal S}(VII_h)$ & VII$_h$ & $k^2\geq 2$ &
	$\ \ \ \ l>\frac{k^2}{4(k^2-2)(4-k^2)}$ for $2<k^2\leq4$ \\ 
		    &&&$l<\frac{k^2}{4(k^2-2)(k^2-4)}$ for $k^2>4$ 
\end{tabular}
\caption{{\em This table lists all of the sinks in the
various Bianchi invariant sets for $2/3<\gamma<2$.  A subset of ${\cal
K}_{\cal M}$ acts as a source for all Bianchi class B models.
$^\dagger$Note: in this case $N_+=0$ (i.e., P$_{\cal S}$=P$_{\cal
S}^\pm$) and in fact corresponds to a Bianchi I model.\label{Table_old_IV}
}}
\end{table}
The function $W_3$ is monotonically decreasing to zero in
$B(I)\backslash S(I)$ and $B(V)\backslash S(V)$.  This implies that
there do not exist any periodic or recurrent orbits in these sets and,
furthermore, the global behaviour of the Bianchi type I and V models
can be determined from the local behaviour of the equilibrium points
in these sets.  We conjecture that there do not exist any periodic or
recurrent orbits in the entire phase space for $\gamma>2/3$, whence it
follows that all global behaviour can be determined from Table
\ref{Table_old_IV}.

We note that a subset of ${\cal K}_{\cal M}$ with
$(1+\Sigma_+)^2>3\tilde\Sigma$, $\Psi>-\sqrt{6}/k$ acts as a source
for all Bianchi class B models.  For $k^2<2$, P$_{\cal S}(I)$ is the
global attractor (sink).  From Table 
\ref{Table_old_IV} 
we see that
there are unique global attractors (both past and future) in all
invariant sets and hence the asymptotic properties are simple to
determine.  The sinks and sources for a particular Bianchi invariant
set, which may appear in that invariant set or on the boundary
corresponding to a (lower-dimensional) specialization of that Bianchi
type, can be easily determined from Table 
\ref{Table_old_IV} 
and Figure 1 which lists the specializations of the Bianchi class B models
\cite{MacCallum1971a}.

 \begin{figure}[h]
  \centering
   \epsfbox{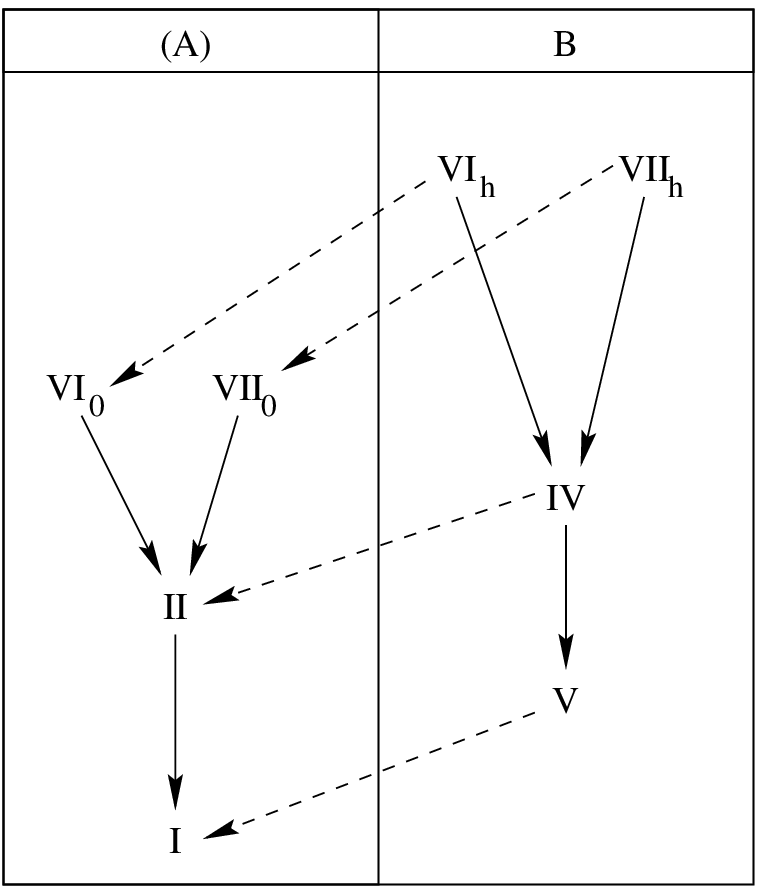}
 \end{figure}
\begin{center}
\noindent Figure 1: {\em Specialization diagram for Bianchi class B models
obtained by letting a non-zero parameter go to zero.  A broken arrow
indicates the group class changes (from B to A).}
\end{center}

\

\

The most general models are those of Bianchi types VI$_h$ and VII$_h$.
The Bianchi type VII$_h$ models are of particular physical interest
since they contain open FRW models as special cases.  From Table 
\ref{Table_old_IV}
and Figure 1 we 
argue
that generically these models (with a scalar
field) isotropize to the future, a result which is of great
significance.  The Bianchi type VI$_h$ models are also of interest
since they contain a class of anisotropic scaling solutions that act
as attractors for an open set of Bianchi type B models.  We note that
generically Bianchi type VI$_h$ models do not isotropize for $k^2\geq
2$.

\section{Intermediate Behaviour and the Invariant Set S(VI$_{\rm h}$)}

It is also of interest to determine the intermediate behaviour of the
models.  In order to do this, we need to investigate the saddles,
determine the dimension of their stable submanifolds, and construct
possible heteroclinic sequences.  This could then be used, in
conjunction with numerical work, to establish the physical properties
of the models.  For example, we could investigate whether {\em
intermediate isotropization} can occur in Bianchi type VII$_h$ models
\cite{WainwrightColeyEllisHancock}.  There are many different cases to
consider depending upon the various bifurcation values and the
particular Bianchi invariant set under investigation.  
As an example,
we shall study the heteroclinic sequences in the
four-dimensional invariant set $S(VI_h)$, because it illustrates the
method and because such a study emphasizes the importance of
anisotropic scaling solutions.


The subspace $S(VI_h)$, which arises from the restrictions $N_+=\Delta=0$ and
\mbox{$3 \Sigma_+^2+l \tilde \Sigma=0$} \cite{WainwrightEllis} is, in
fact, the class of diagonal Bianchi VI$_h$ models and is four--dimensional 
(and was shown in \cite{ColeyIbanezOlasagasti} to
illustrate the existence and importance of the anisotropic scaling
solutions). From the above restrictions, the system of equations
(\ref{DE_1})-(\ref{DE_7}) now reduce to:
\begin{eqnarray}
\label{DE_6_1}
& \Sigma_+'=(q-2) \Sigma_+ +\frac{2}{3}l \tilde A  &  \\
& \tilde A'=2(q+2 \Sigma_+) \tilde A & \\
& \Psi'=(q-2) \Psi-{1 \over 2}\sqrt{6}k \Phi^2 & \\
\label{DE_6_4}
& \Phi'=(q+1+{1 \over 2}\sqrt{6}k \Psi) \Phi& 
\end{eqnarray}
We note that the constraint equation (\ref{constraint1}) is automatically 
satisfied.

The equilibrium points of the system are a subset of the ones
presented in section III and we present those that belong to this subspace and
their corresponding eigenvalues in Table VI:
\noindent
\begin{center}
\begin{tabular}{|l|c|ll|l|} \hline
Eqm point/set &Eigenvalues& Stability & & Conditions \\
\hline \hline
${\cal L}^\pm_k$ & $  {6 \over l-3} \q {6 \over l-3} \q 3{l-1 \over l-3}
     \q  {3 \over 2}{(3 \gamma-2)+l(2-\gamma) \over l-3}    $ & 
	$s^2 \leq 3\gamma-2:$ & $ W^s:\{\Phi=0\} $ & \\ &
         & $s^2 > 3\gamma-2:$  & $W^s:\{\Phi=\Omega=0\}$ &  \\ 
\hline
$\cal K$$_{\cal M}$ &  $0 \q 3(2-\gamma) \q 4(1+\Sigma_+) $
      & $k^2<6$ & $W^s:\{ \emptyset\}$ & \\ &  $3+{\sqrt{6} \over 2}k \sqrt{1+{3-l \over l}\Sigma_+^2}$ 
      &$k^2\geq 6$ and $\Sigma_0^2>\frac{6}{k^2}$ &$W^s:\{ \emptyset \}$ & \\ &
      &$k^2\geq 6$ and $\Sigma_0^2<\frac{6}{k^2}$ &$W^s:\{ \Omega=K=0\}$ & \\
\hline
$P(I)$ & $-{3 \over 2}(2-\gamma) \q -{3 \over 2}(2-\gamma) \q
          3 \gamma-2 \q  {3 \over 2}\gamma $ & & $W^s:\{\Phi=K=0\} $   & \\
\hline
$P(VI_h)$ & $ {3 \over 2}\gamma \q -{3 \over 2}(2-\gamma) \q
        -{3 \over 4}(2-\gamma)(1 \pm \sqrt{1-r^2}) $ 
 	& & $W^s:\{ \Phi=0 \}$ &  $s^2>3\gamma-2$\\
\hline
$P_{\cal S}(I)$ & ${k^2-6 \over 2} \q {k^2-6 \over 2} \q
         k^2-2 \q k^2-3 \gamma$
	& $k^2\leq 2$ & $W^s:\{S(VI_h) \}$ & $k^2<6$\\ &
                &$2<k^2<3 \gamma$ & $W^s:\{ K=0 \}$ &  \\ &
                &$k^2>3 \gamma$  &  $W^s:\{ K=\Omega=0 \}$ & \\ 
\hline
$P_{\cal S}(VI_h)$ & $6\frac{k^2-2l}{[k^2(l-3)+4l]} \q
       -3\gamma-6\frac{k^2(1-l)}{[k^2(l-3)+4l]}$
	& $k^2< 3 \gamma$ or $s^2<s_0^2$ 
         & $  W^s:\{ S(VI_h) \}$  &  $k^2>2 $\\ & $3\frac{(k^2-2l)\pm\sqrt{(k^2-2l)[(k^2-2l)-4(1-l)(k^2-2)]}}{[k^2(l-
	3)+4l]} $
         & $s^2>s_0^2$ & 
         $  W^s:\{ \Omega=0 \}$ & \\ 
\hline
${\cal F}_S(I)$ & $-{3 \over 2}(2-\gamma) \q 3\gamma-2$ 
	& & $  W^s:\{ K=0 \}$ & $k^2>3 \gamma$ \\ & $ -{3 \over
	4}(2-\gamma) \pm \sqrt{(2-\gamma)(2-9 \gamma+24 \gamma^2/k^2)}$ &&& \\
\hline
${\cal A}_S(VI_h)$ &  $ -\frac{3}{4}(2-\gamma) \pm\sqrt{a\pm \sqrt{b}}$
	& &  $  W^s:\{ S(VI_h) \}$ 
    & $s^2>s_0^2$  \\
\hline
\end{tabular}
\end{center}

\

\noindent{Table VI: {\em This table lists all equilibrium points of the
system (\ref{DE_6_1})-(\ref{DE_6_4}) and gives the corresponding stable (sub)manifold, $W^s$. We also give the conditions on the
parameters for their existence.  We note that $s^2 \equiv {4l \over
l-3}$, $s_0^2 \equiv \frac{k^2(3\gamma-2)}{(k^2-3\gamma)}$, $\Sigma_0^2 \equiv
1-\tilde \Sigma-\Sigma_+^2=1-{l-3 \over l} \Sigma_+^2$, $r^2\equiv
2(3\gamma-2)(1-l_c/l)$, $q^2\equiv 2r^2/(2-\gamma)$ and $l_c\equiv
-(3\gamma-2)/(2-\gamma)$.  Note that the four eigenvalues for ${\cal
A}_S(VI_h)$ are the same as the last four eigenvalues in
Eq. (\ref{bugger}), but are listed here in ``short form'' for compact
notation.}}

\vspace{0.5cm}

>From the eigenvalues in Table V we can study the stability of the equilibrium
points and the qualitative behaviour in this four-dimensional subspace. The
bifurcation values leading to a change in the asymptotic behaviour are:
$k^2=2, \q k^2=6, \q k^2=3 \gamma, \q s^2=3\gamma-2, \q
s^2=k^2(3\gamma-2)/(k^2-3\gamma)$. 

We can see that the possible future asymptotic behaviour is given
by points containing a scalar field, either alone or together with a 
perfect fluid component in the case of the anisotropic scaling solution. 
These possible future asymptotics are:  $P_{\cal S}(I)$, $P_{\cal
S}(VI_h)$ and ${\cal A}_S(VI_h)$.

\subsection{Heteroclinic Sequences}

The stable and unstable manifolds of the saddle equilibrium points
provide a skeleton of special orbits that play a significant r\^{o}le
in determining the dynamics of the models (and in particular, their
intermediate states) \cite{WainwrightColeyEllisHancock}.  A (finite)
heteroclinic sequence is a set of equilibrium points $E_0, E_1,
\ldots, E_n$, where $E_0$ is a local source, $E_n$ is a local sink and
the rest are saddles, such that there is a heteroclinic orbit which
joins $E_{i-1}$ to $E_i$ for each $i=1,\ldots,n$
\cite{WainwrightEllis}.  Below are the heteroclinic orbits in the
class of $S(VI_h)$ models for different parameter ranges.  Equilibrium
points in parentheses stand for optional intermediate points.  We
recall that $s^2 \equiv {4l \over l-3}$ and $s_0^2\equiv
\frac{k^2(3\gamma-2)}{(k^2-3\gamma)}$.  Finally, when $k^2>6$ the
Kasner-like ring splits into $\ptwo[A]$ (sources) and $\ptwo[B]$
(saddles.)


\

\begin{minipage}[t]{17cm}
\sloppy
\parbox[t]{2cm}{\sloppy\fbox{$k^2<2<3\gamma$}} \hfill
\parbox[t]{6cm}{i) \underline{$s^2< 3\gamma-2$}: \\
$\ptwo \rightarrow \pone \rightarrow \pthree$ \\
$\ptwo \rightarrow \pfive \rightarrow \pone \rightarrow \pthree$}\hfill
\parbox[t]{6cm}{ii) \underline{$3\gamma-2< s^2$} \\
$\ptwo \rightarrow \pone \rightarrow \psix \rightarrow \pthree$\\
$\ptwo \rightarrow \pfive \rightarrow \psix \rightarrow \pthree$} \\
\mbox{}
\end{minipage}

\vspace{1cm}

\begin{minipage}[t]{17cm}
\sloppy
\parbox[t]{2cm}{\sloppy\fbox{$2<k^2<3\gamma$}} \hfill
\parbox[t]{6cm}{i) \underline{ $s^2< 3\gamma-2$}: \\
\sloppy $\ptwo \rightarrow \pfive \rightarrow \pthree \rightarrow \pfour$\\
$\ptwo \rightarrow \pfive \rightarrow \pone \rightarrow \pfour$}\hfill
\parbox[t]{6cm}{ii) \underline{$3\gamma-2< s^2$}:\\
\sloppy $\ptwo \rightarrow \pone \rightarrow (\psix) \rightarrow \pfour$ \\
$\ptwo \rightarrow \pfive \rightarrow  \psix \rightarrow \pfour$ \\
$\ptwo \rightarrow \pfive \rightarrow  \pthree \rightarrow \pfour$} \\
\mbox{}
\end{minipage}

\vspace{1cm}

\begin{minipage}[t]{17cm}
\sloppy
\parbox[t]{2cm}{\sloppy\fbox{$2<3\gamma<k^2< 6$}} \hfill
\parbox[t]{6cm}{i) \underline{$s^2< 3\gamma-2< s_0^2$}: \\
\sloppy $\ptwo \rightarrow \pthree \rightarrow (\pseven) \rightarrow \pfour$ \\
$\ptwo \rightarrow (\pfive) \rightarrow \pone \rightarrow \pfour$ \\
$\ptwo \rightarrow \pfive \rightarrow \pseven \rightarrow \pfour$}\hfill
\parbox[t]{6cm}{ii) \underline{$3\gamma-2< s^2< s_0^2$}: \\
\sloppy $\ptwo \rightarrow \pone \rightarrow (\psix) \rightarrow \pfour$ \\
$\ptwo \rightarrow \pthree \rightarrow (\pseven) \rightarrow \pfour$\\
$\ptwo \rightarrow \pfive \rightarrow \psix \rightarrow \pfour$\\
$\ptwo \rightarrow \pfive \rightarrow \pseven \rightarrow \pfour$} \\
\mbox{} \\ 
\parbox[t]{2cm}{\hfill} \hfill
\parbox[t]{8cm}{iii) \underline{$3\gamma-2< s_0^2< s^2$}:\\
\sloppy $\ptwo \rightarrow \pone \rightarrow \pfour \rightarrow \peight$\\
$\ptwo \rightarrow \pone \rightarrow \psix \rightarrow \peight$\\
$\ptwo \rightarrow \pthree \rightarrow \pfour \rightarrow \peight$\\
$\ptwo \rightarrow \pthree \rightarrow \pseven \rightarrow \peight$\\
$\ptwo \rightarrow \pfive \rightarrow \psix \rightarrow \peight$\\
$\ptwo \rightarrow \pfive \rightarrow \pseven \rightarrow \peight$}\hfill
\parbox[t]{4cm}{\hfill}\\
\mbox{}
\end{minipage}

\vspace{1cm}

\begin{minipage}[t]{17cm}
\sloppy
\parbox[t]{2cm}{\sloppy\fbox{$3\gamma<6<k^2$} }\hfill
\parbox[t]{10cm}{i) \underline{$s^2< 3\gamma-2< s_0^2$}:\\
\sloppy $\ptwo[A] \rightarrow (\ptwo[B]) \rightarrow (\pfive) \rightarrow \pone \rightarrow \pfour$ \\
$\ptwo[A]  \rightarrow (\ptwo[B]) \rightarrow \pfive \rightarrow \pseven \rightarrow \pfour$ \\ } \hfill \parbox[t]{2cm}{}\\
\parbox[t]{2cm}{\hfill} \hfill
\parbox[t]{10cm}{ii) \underline{$3\gamma-2< s^2< s_0^2$}:\\
\sloppy $\ptwo[A] \rightarrow (\ptwo[B]) \rightarrow \pone \rightarrow (\psix) \rightarrow \pfour$\\
$\ptwo[A] \rightarrow (\ptwo[B]) \rightarrow \pfive \rightarrow \psix \rightarrow \pfour$\\
$\ptwo[A] \rightarrow (\ptwo[B]) \rightarrow \pfive \rightarrow \pseven \rightarrow \pfour$ \\ } \hfill \parbox[t]{2cm}{}\\
\parbox[t]{2cm}{\hfill} \hfill
\parbox[t]{10cm}{iii) \underline{$3\gamma-2< s_0^2< s^2$}:\\
\sloppy $\ptwo[A] \rightarrow (\ptwo[B]) \rightarrow \pone \rightarrow \pfour \rightarrow \peight$\\
$\ptwo[A] \rightarrow (\ptwo[B]) \rightarrow \pone \rightarrow \psix \rightarrow \peight$\\
$\ptwo[A] \rightarrow (\ptwo[B]) \rightarrow \pfive \rightarrow \psix \rightarrow \peight$\\
$\ptwo[A] \rightarrow (\ptwo[B]) \rightarrow \pfive \rightarrow \pseven \rightarrow \peight$} \hfill \parbox[t]{2cm}{}\\
\mbox{}
\end{minipage}

\

\begin{center}
\setlength{\unitlength}{1cm}
\begin{picture}(15,8)
\thicklines
\put(0.875,3){\makebox(2,2){$\ptwo$}}
\put(4.625,0.333){\makebox(2,2){$\pfive$}}
\put(4.625,3){\makebox(2,2){$\pthree$}}
\put(4.625,5.667){\makebox(2,2){$\pone$}}
\put(8.375,0.333){\makebox(2,2){$\pseven$}}
\put(8.375,3){\makebox(2,2){$\psix$}}
\put(8.375,5.667){\makebox(2,2){$\pfour$}}
\put(12.125,3){\makebox(2,2){$\peight$}}
\put(1.875,4.5){\line(0,1){2.167}} \put(1.875,6.667) {\vector(1,0){3.125}}
\put(2.375,4) {\vector(1,0){2.625}}
\put(1.875,3.5){\line(0,-1){2.167}} \put(1.875,1.333) {\vector(1,0){3.125}}
\put(6.25,6.667){\vector(1,0){2.0}}
\put(6.25,1.333){\vector(1,0){2.0}}
\put(10.5,6.667){\line(1,0){2.625}} \put(13.125,6.667){\vector(0,-1){2.167}}
\put(10.5,4){\vector(1,0){1.625}}
\put(10.5,1.333){\line(1,0){2.625}} \put(13.125,1.333){\vector(0,1){2.167}}
\put(6.25,4.5){\vector(4,3){2}}
\put(6.25,3.5){\vector(4,-3){2}}
\put(6.25,6){\vector(4,-3){2}}
\put(6.25,2){\vector(4,3){2}}
\end{picture}
\end{center}
\noindent{Figure 2.  {\em Heteroclinic sequences in $S(VI_h)$ for
$s<3\gamma<k^2<6$ and $3\gamma-2<s_0^2<s^2$, indicating the skeleton
of orbits defined by the stable and unstable manifolds of the saddle
points.  Note that the anisotropic scaling solution ${\cal A}_S(VI_h)$
is a stable attractor.}}

Figure 2 can be used to construct sequences of orbits joining
equilibrium points starting at the past attractor $\ptwo$ and ending
at the future attractor ${\cal A}_S(VI_h)$.  For each sequence there
will be a family of orbits that shadow this sequence in the state
space.

\section{Conclusion}

In this paper we have discussed the qualitative properties of Bianchi
type B cosmological models containing a barotropic fluid and a scalar
field with an exponential potential.  The most general models are
those of type VI$_h$, which include the anisotropic scaling solutions,
and those of type VII$_h$, which include the open FRW models.  

In cases in which we have been able to find monotone functions we have
been able to prove global results.  Otherwise, based on the local
analysis of the stability of equilibrium points and the dynamics on
the boundaries of the appropriate state space, we have presented
plausible global results (this is similar to the analysis of perfect
fluid models in \cite{WainwrightEllis} and \cite{HewittWainwright} in
which no monotone functions were found in the Bianchi type VI and VII
invariant sets).  In all cases, however, our results are further
justified by numerical experimentation.  

Let us summarize the main
results:

\begin{itemize}
\item All models with $k^2<2$ asymptote toward the flat FRW power-law
inflationary model \cite{R3,ColeyIbanezVanDenHoogen},
corresponding to the global attractor $P_{\cal S}(I)$, at late times;
i.e., all such models isotropize and inflate to the future.

\item $F_{\cal S}(I)$ is a saddle and hence the flat FRW scaling
solutions \cite{Wetterich,Wands} do not act as late-time
attractors in general \cite{BillyardColeyVanDenHoogen}.

\item A subset of ${\cal K}_{\cal M}$ acts as a source for all Bianchi
type B models; hence all models are asymptotic in the past to a massless
scalar field analogue of the Jacobs anisotropic Bianchi I solutions.

\item For $k^2\geq 2$, Bianchi type VII$_h$ models generically
asymptote towards an open FRW scalar field model, represented by one
of the local sinks $P_{\cal S}(V)$ or $P_{\cal S}^\pm(VII_h)$, and
hence isotropize to the future.

\item For $k^2\geq 2$, Bianchi type VI$_h$ models generically
asymptote towards either an anisotropic scalar field analogue of the Collins
solution \cite{Collins}, an anisotropic vacuum solution (with no
scalar field) or an anisotropic scaling solution
\cite{ColeyIbanezOlasagasti}, corresponding to the local sinks
$P_{\cal S}^\pm(VI_h)$, ${\cal L}_k(VI_h)$ or ${\cal A}_{\cal
S}(VI_h)$, respectively (depending on the values of a given model's
parameters - see table IV for details).  These models do not generally
isotropize.

\item In particular, the equilibrium point ${\cal A}_{\cal S}(VI_h)$
is a local attractor in the Bianchi VI$_h$ invariant set and hence
there is an open set of Bianchi type B models containing a perfect
fluid and a scalar field with exponential potential which asymptote
toward a corresponding anisotropic scaling solution at late times.
\end{itemize}

We should stress that our analysis and results are applicable to a
variety of other cosmological models in, for example, scalar-tensor
theories of gravity (which are formally equivalent to general
relativity containing a scalar field with an exponential potential)
\cite{BillyardColeyIbanez,Barrow1994a,Liddle1992a,Mimoso1995a},
theories with multiple scalar fields with exponential potentials
\cite{Liddle1998a} and string theory \cite{BCL,Kaloper1998b}.

In future work we shall study spatially homogeneous models with
positive spatial curvature \cite{Coley1999a} and Bianchi models of
type A \cite{Ibanez1999a}.  However, our ultimate goal is to extend
the techniques used in this paper and study the more interesting (and
physically more relevant) case of spatially inhomogeneous models.

\

\

\centerline{\bf Acknowledgments}

APB is supported by Dalhousie University. AAC and RJvdH is supported by the
Natural Sciences and Engineering Research Council of Canada (NSERC).
JI and IO is supported by the CICYT (Spain) Grant
PB96-0250.


\end{document}